\documentclass[aps,prl,floatfix,reprint,twocolumn,amsmath,superscriptaddress,amssymb,showpacs]{revtex4-2}
\usepackage{amsmath,braket,amssymb}
\usepackage[final]{graphicx}
\usepackage{subfigure}
\usepackage{xcolor}
\usepackage[english]{babel}
\usepackage[utf8]{inputenc}
\usepackage{color}
\usepackage{mathtools}

\usepackage[colorlinks,citecolor=darkBlue,linkcolor=darkBlue,
urlcolor=blue,hyperindex]{hyperref}

\usepackage[normalem]{ulem}
\normalfont

\definecolor{forestgreen}{rgb}{0.08, 0.4, 0.13}
\definecolor{darkBlue}{rgb}{0.08, 0.13, 0.4}

\begin{document}

\title{Entanglement and Correlation Spreading in non-Hermitian Spin Chains}

\author{Xhek Turkeshi}
\author{Marco Schir\'o}
\affiliation{JEIP, USR 3573 CNRS, Coll\`{e}ge de France, PSL Research University, 11 Place Marcelin Berthelot, 75321 Paris Cedex 05, France}

\begin{abstract}
Non-Hermitian quantum many-body systems are attracting widespread interest for their exotic properties, including unconventional quantum criticality and topology.
Here we study how quantum information and correlations spread under a quantum quench generated by a prototypical non-Hermitian spin chain. Using the mapping to fermions we solve exactly the problem and compute the entanglement entropy and the correlation dynamics in the thermodynamic limit. 
Depending on the quench parameters, we identify two dynamical phases. One is characterized by rapidly saturating entanglement and correlations. The other instead presents a logarithmic growth in time, and correlations spreading faster than the Lieb-Robinson bound, with collapses and revivals giving rise to a modulated light-cone structure. Here, in the long-time limit, we compute analytically the entanglement entropy that we show to scale logarithmically with the size of the cut, with an effective central charge that we obtain in closed form. Our results provide an example of an exactly solvable non-Hermitian many-body problem that shows rich physics including entanglement and spectral transitions. 
\end{abstract}

\maketitle

Unitarity is a fundamental property of quantum mechanics~\cite{Shankar:102017} which puts constraints on how quantum information and correlations spread in a many-body system, as it has been understood in recent years through the study of nonequilibrium dynamics of isolated quantum many-body systems~\cite{lieb72thefinite,calabrese2005evolution,polkolnikov2011colloquium,cheneau2012lightcone,Gogolin_2016}.
In fact, unitarity reflects core aspects of quantum mechanics, such as probability conservation and the linearity of the Sch\"odinger equation. But what happens to correlation and entanglement spreading when the dynamics of the system is non-unitary?
These questions originate from the interest around the non-unitary quantum mechanics generated by non-Hermitian Hamiltonians, which has a long tradition dating back to works on noninteracting electronic disordered systems~\cite{hatano1996localization,efetov1997directed}, non-unitary conformal field theories~\cite{cardy1985conformal,fisher1978yanglee} and it has been recently revived~\cite{ashida2020non}.
A typical setup in which non-Hermitian evolution naturally appears is in the context of open quantum systems and quantum optics, where it stems from the coupling to the measurement apparatus (backaction) and subsequent post-selection of quantum trajectories where no jump process occurs~\cite{Wiseman2009}.

For few-body non-Hermitian systems the focus has been mainly on spectral properties, in particular the existence of exceptional points~\cite{Heiss_2012}, parity-time reversal symmetry~\cite{bender1998real,bender2015pt} and their experimental realization in optical systems with gain and losses~\cite{guo2009observation,ruter2010observation}.
In the many-body domain, non-Hermitian systems host novel phenomena compared to the Hermitian counterparts, including unconventional critical phenomena ~\cite{ashida2017parity} and quantum impurity physics~\cite{nakagawa2018nonhermitian,lourenco2018kondo,yoshimura2020nonhermitian}, unusual nonequilibrium dynamics after quantum quenches for PT-symmetric Luttinger Liquids~\cite{dora2020quantum,moca2021universal}, exotic entanglement~\cite{couvreur2017entanglement,herviou2018entanglement,chang2020entanglement}, and topological properties~\cite{gong2018topological}. For PT-symmetric free-fermionic systems it has been shown that entanglement grows linearly in time and saturates to a volume law~\cite{ashida2018full,bacsi2021dynamics,dora2021correlations}, similarly to their Hermitian counterpart. Yet, the dynamics of entanglement and correlations after a quench in a non-Hermitian quantum many-body system is largely an unexplored field. Recently, hints that a richer pattern of entanglement and correlation dynamics, unique to the non-Hermitian setting, may arise have emerged in the study of continuously monitored quantum many-body systems~\cite{alberton2020entanglement,buchold2021effective,turkeshi2021measurementinduced,turkeshi2021entanglement,fuji2020measurementinduced}. It is therefore desirable to study the non-unitary evolution of quantum correlations of an analytically treatable quantum system, and to provide exact results on this framework.

In this work we study the quench dynamics of a prototypical non-Hermitian spin chain: the anisotropic XY model with a complex transverse field. Using the Jordan-Wigner transformation, we compute exactly the dynamics of entanglement entropy and spin-spin correlations. 
Similar quantum spin chains with complex parameters have been introduced before~\cite{lee2014heralded,Pi_2021}, however its quantum dynamics has not been discussed so far, in particular concerning entanglement and correlation spreading for which we are going to provide exact results in the thermodynamic limit. We show that the former grows logarithmically slow in time up to a critical value of the transverse field and in this phase, which is characterized by a gapless dispersion of decay modes, the correlations spread anomalously faster than the Lieb-Robinson bound. Above a critical value of the transverse field the system enters a phase with gapped decay modes, where entanglement saturates quickly  to an area-law, and spin-spin correlations are exponentially localized. 
We then focus on the long time limit and compute analytically the leading order entanglement entropy in the thermodynamics limit. We show that the system undergoes an entanglement transition, which mirrors and is driven by the spectral properties of the system~\cite{biella2021manybody,gopalakrishnan2021entanglement,kells2021topological}. 
\begin{figure*}[t]
    \centering
    \includegraphics[width=\textwidth]{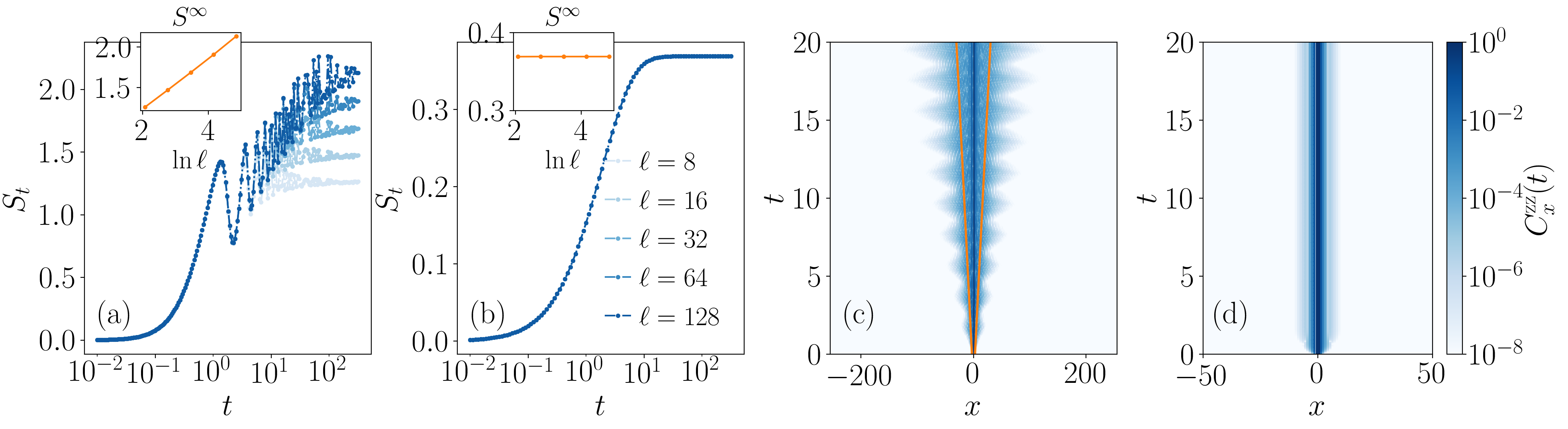}
    \caption{\label{fig:czz} (a-b) Entanglement dynamics in the anisotropic non-Hermitian XY chain at fixed $\kappa=0.85$, $h=0.3$. For $\gamma=1.2$ (a) the growth is logarithmic in time and saturates to a value $S^{\infty}$ which scales logarithmically with the size of the cut $\ell$ (inset), while for $\gamma=4.2$ (b) the entanglement quickly saturates to a value independent from $\ell$. Spreading of the correlation $C^\mathrm{zz}_x(t)$ as a function of spacetime for $\kappa=0.85$, $h=0.3$, and $\gamma=0.8$, $\gamma=6.2$ for respectively the panel (c) and (d). In the first case the spreading is faster than the Lieb-Robinson bound of the unitary dynamics generated by the $H_\mathrm{XY}(h,\kappa)$ (continuous orange line). For panel (d), corresponding to a quench to large $\gamma$, the correlations are sizeable only in a narrow range around the origin. }
\end{figure*}

\textit{Non-Hermitian quantum quench ---} We consider the following non-Hermitian quantum quench protocol, in which the system at $t=0$ is initialized in the ground state $|\Psi_0\rangle$  of the anisotropic XY chain with Hamiltonian
\begin{align}
H_\mathrm{XY} = -\sum_{i=1}^L \left[\frac{1+\kappa}{2}\hat\sigma^x_i \hat\sigma^x_{i+1}+ \frac{1-\kappa}{2}\hat\sigma^y_i \hat\sigma^y_{i+1} + h \hat\sigma^z_i\right]
\end{align}
with periodic boundary conditions and initial values ${h=h_0}$ and ${\kappa=\kappa_0}$ and for $t>0$ is let evolve under the action of the non-Hermitian anisotropic XY chain
\begin{align}
  	|\Psi(t)\rangle &= \frac{e^{-i \hat{H}_\mathrm{eff} t}|\Psi_0\rangle}{||e^{-i \hat{H}_\mathrm{eff} t}|\Psi_0\rangle||},\  
   	H_\mathrm{eff} = H_\mathrm{XY} - i\frac{\gamma}{4}\sum_{i=1}^L \sigma^z_i .\label{eq:main_quench} 
\end{align}
corresponding to a sudden switching of the dissipative non-Hermitian coupling $\gamma$ and a quench of the transverse field and asymmetry parameter $h,\kappa$. The dynamics Eq.~\eqref{eq:main_quench} naturally arises as the no-click limit of the stochstic quantum jump trajectories~\cite{dalibard1992wavefunction,daley2014quantum} when $\hat{n}_i=(1+\hat{\sigma}^z_i)/2$ is monitored~\cite{SM} and in this context $\gamma$ plays the role of measurement rate controlling the role of the back-action term. As such we will refer to this parameter either as dissipation or measurement rate.

Without loss of generality, we consider an initial state with well-defined parity $\prod_{i} \sigma^z_i$. The time-evolution Eq.~(\ref{eq:main_quench}) can be exactly computed once the model is mapped to free fermions through Jordan-Wigner transformation, followed by Fourier transform into momentum space  and a complex Bogolubov rotation~\cite{SM}. The  state at time $t$ is
\begin{equation}
	|\Psi_t\rangle = \prod_{k\in \mathcal{K}} \left(\frac{u_k(t) + v_k(t) \hat{c}^\dagger_k \hat{c}^\dagger_{-k}}{\sqrt{|u_k(t)|^2+|v_k(t)|^2}} \right)|0\rangle\;, \label{eq:main_exactsolution}
\end{equation}
where the momenta are restricted to the set $\mathcal{K}=\{ (2n-1)\pi/L | n=0,1,\dots,L/2\}$, fixed by the parity of the initial state and by the boundary conditions. The coefficients
\begin{equation}
    \begin{split}
	u_k(t) &= \cos(\theta_k^0)\cos(\Lambda_k t) -i \cos(\Phi_k)\sin(\Lambda_k t) \\
	v_k(t) &= \sin(\theta_k^0)\cos(\Lambda_k t) +i \sin(\Phi_k)\sin(\Lambda_k t)
    \end{split}\label{eq:main_wvsolk}
\end{equation}
are written in terms of the Bogolubov angles $\Phi_k\equiv \theta_k - \theta_k^0$~\footnote{These angles are fixed by the initial condition and by the quench parameters through $\cos(\theta^0_k) = u_k(0)$ and $\cos(\theta_k) = E_k/\Lambda_k$.} and (complex) quasiparticle spectrum  $\Lambda_k= E_k + i \Gamma_k$, which is given by
\begin{align}
\Lambda_k =  \pm\sqrt{\left(2\cos k+2 h+i\gamma/2\right)^2+4\kappa^2\sin^2 k}\, .
\end{align}
The convention on the sign of $\Lambda_k$ can be independently chosen for every momentum $k$~\cite{lee2014heralded}, and in particular is fixed in such a way that $\Gamma_k\le 0$.
Therefore, the quasiparticle spectrum $\Lambda_k$ governs the dynamical and stationary properties of the system. In particular, the spectral transition occurring at a critical value of the dissipation rate $\gamma$~\cite{biella2021manybody,turkeshi2021entanglement} translates to transition in the correlations and in the entanglement. 
In fact, for ${\gamma<\gamma_c(h,\kappa)}$ the imaginary part of the spectrum vanishes linearly at two $k$ points $k=\pm k^*$, while for $\gamma>\gamma_c$ the spectrum of decay rates is gapped (see Ref.~\cite{SM}). The existence of these two long-lived modes has crucial consequences for the dynamics of entanglement and correlations, as we are going to discuss next.
Since the order of the limits plays a fundamental role, in this Letter we consider the non-trivial ${\gamma\to 0}$ (${\kappa \to 0}$) by first fixing ${\gamma=\epsilon}$ (${\kappa = \epsilon}$), computing the evolution Eq.~\eqref{eq:main_exactsolution}, and interpolating ${\epsilon \to 0}$~\footnote{The \textit{ab-initio} $\gamma=0$ and $\kappa=0$ correspond, respectively, to the unitary evolution in Ref.~\cite{fagotti2008evolution,calabrese2005evolution}, and to the free fermion unitary evolution~\cite{calabrese2009entanglement}, as the imaginary part of the non-Hermitian Hamiltonian commutes with $H_\mathrm{XY}$ and simplifies in the ratio with the norm (cfr. Eq.~\eqref{eq:main_quench})}. This has relevant consequences, as we discuss below.

For later purpose it is convenient to introduce the functions $\Pi_{\alpha}(k,t)$ which encode certain combinations of the coefficients $u_k(t),v_k(t)$ that we collect in a two-dimensional spinor $\overline{\varphi}(t)_k=\left( \overline{u}(t)_k\,,\, \overline{v}(t)_k\right)$
\begin{align}
	\Pi_{\alpha}^t(k) &= \frac{\overline{\varphi}_k(t) R_x(\pi/2)\tau_{\alpha}R_x(\pi/2)^\dagger\varphi_k(t)}{\overline{\varphi}_k(t)\varphi_k(t)}	
	\label{eq:Pi_alpha}
\end{align}
where $\tau_{\alpha=x,y,z}$ are Pauli matrices, $R_x(\theta) = \exp(-i \tau_x \theta/2)$ and the bar denotes the complex conjugation.

\textit{Dynamics of entanglement and correlation ---}  
We start discussing the dynamics of the entanglement entropy. Throughout this work we consider a bipartition $X\cup X_c$ where $X$ is a connected interval of length $\ell$ and $X_c$ is the complementary.  The entanglement entropy is defined as $S_t(\ell)\equiv-\mathrm{tr}(\rho_X \ln \rho_X)$, where $\rho_X(t) = \mathrm{tr}_{X_c}|\Psi(t)\rangle\langle \Psi(t)|$ is the reduced density matrix. 
Due to the Gaussianity of the state, this quantity is fully characterized by the two-point fermionic correlations~\cite{vidal2003entanglement,jin2005entanglement,peschel2004on}.
In terms of the Majorana fermions $\hat{a}_{2l-1} = \hat{c}^\dagger_l + \hat{c}_l $ and $\hat{a}_{2l}  = -i (\hat{c}^\dagger_l - \hat{c}_l)$, the latter are given by 
\begin{align}
	2\mathbb{A}_{mn} \equiv \langle \Psi_t|\hat{a}_m\hat{a}_n|\Psi_t\rangle = \delta_{mn} - \Gamma^{\ell}_{mn}(t)\label{eq:aaaa}\;,
\end{align}
where the block Toeplitz matrix $\Gamma^\ell$ is generated by the symbol Eq.~\eqref{eq:Pi_alpha}
\begin{equation}
	\Gamma^\ell_{mn}(t)  = \int_{-\pi}^\pi \frac{dk}{2\pi} e^{-i k (m-n)} \vec{\Pi}^t(k)\cdot \vec{\sigma}\;.\label{eq:toepliz}
\end{equation} 
Using Wick's theorem, the entanglement entropy reduces to ${S_t(\ell) = -\mathrm{tr}(\mathbb{A}\ln \mathbb{A})}$, which requires the evaluation of the spectrum of $\Gamma^\ell$ (cfr. Eq.~\eqref{eq:aaaa}).
Numerically, this can be achieved efficiently, as the complexity scales polynomially with the system size. 
Our data distinguished between two regimes, as exemplified in Fig.~\ref{fig:czz}(a-b). For $\gamma<\gamma_c$, after an initial transient time, the entanglement entropy exhibits a logarithmic growth dressed by damped oscillations (Fig.~\ref{fig:czz}(a)), and saturates to a stationary value which scales logarithmically with the partition size $\ell$ (inset). On the other hand, for $\gamma \ge \gamma_c$, the entanglement saturates quickly to a value independent of $\ell$ (Fig.~\ref{fig:czz}(b) and inset). 
We emphasize that these results are remarkably different compared to unitary many-body systems, \textit{e.g.} for the quench of the isolated XY chain, where the entanglement grows linearly in time~\cite{fagotti2008evolution} and saturates to a volume law~\cite{calabrese2005evolution}. (We note that similar numerical and analytical evidence are also present for integrable systems and random unitary evolution~\cite{alba2017entanglement,nahum2017quantum}). \textit{En passant}, we note that the absence of linear entanglement growth and volume law scaling in the present non-Hermitian setting can be understood by means of the Szeg\"o theorem~\cite{calabrese2005evolution} and its generalization~\cite{fagotti2008evolution}, using the fact that $\det(\vec{\Pi}^t(k)\cdot \vec{\Pi}^t(k)) = 1$ for any $k$, $t$ (see Ref.~\cite{SM} for an in-depth discussion).  Quite interestingly our results also differ from other non-interacting non-Hermitian systems recently studied~\cite{bacsi2021dynamics,dora2021correlations} which displayed linear growth and volume-law scaling of the entanglement entropy. We trace back these differences to the spectral properties of our non-Hermitian spin chain undergoing a subradiance transition from a gapless spectrum of decay modes to a gapped one at $\gamma_c$, as we are going to show explicitly in the following.  For $\gamma<\gamma_c$ therefore the system is in a critical non-Hermitian phase, reminiscent of non-unitary CFT, which results in power-law correlations~\cite{lee2014heralded} and logarithmic scaling of entanglement.

These differences are also present in the spreading of the correlations, as encoded in the connected spin-spin correlation function
\begin{align}
    C^\mathrm{zz}_x(t) &= \langle \Psi_t| \hat\sigma^z_0\hat\sigma^z_x |\Psi_t\rangle -  \langle \Psi_t| \hat\sigma^z_0|\Psi_t\rangle\langle \Psi_t|\hat\sigma^z_x |\Psi_t\rangle 
\end{align}
which can be evaluated in closed form after simple manipulations with the Jordan-Wigner transformation and using the Wick's theorem~\cite{lee2014heralded,SM}.
For unitary systems, this quantity is bound by the Lieb-Robinson bound~\cite{lieb72thefinite,igloi2000longrange,calabrese2007quantum}, which guarantees a maximum sound velocity. Here instead we see that for small values of $\gamma$ correlations propagate outside the light-cone~\cite{ashida2018full,dora2020quantum}, which is blurred and acquire a modulated \emph{crocodile-tail} structure with collapses and revivals which are unique to non-Hermitian systems.  On the other hand for large quenches correlations decays fast and the light cone is much reduced.

\begin{figure}[t]
    \centering
    \includegraphics[width=\columnwidth]{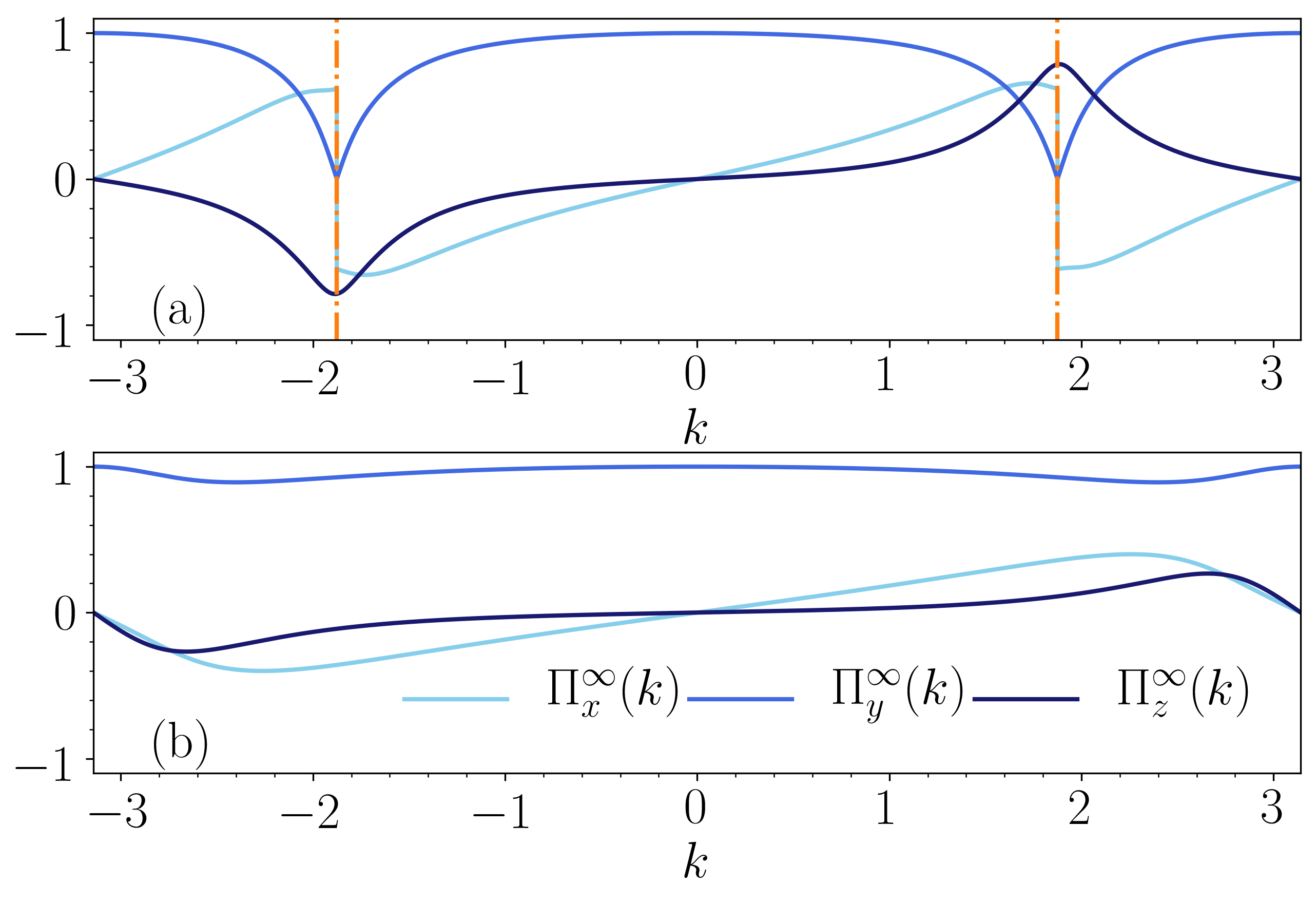}
    \caption{\label{fig:discontinuit} Stationary-state correlators $ \vec{\Pi}^{\infty}(k)$ in the gapless  ($\gamma<\gamma_c$, top) and gapped ($\gamma>\gamma_c$, bottom) phases. We see that in the former case $\Pi_z^{\infty}(k)$ displays a sharp jump discontinuity at $k=\pm k^*$, reminiscent of critical one dimensional fermions, while the other components remain smooth. As we show in the text, this feature is responsible for the logarithmic scaling of the entanglement entropy. On the other hand above $\gamma_c$ we see that $ \vec{\Pi}^{\infty}(k)$ is smooth and featureless.}
\end{figure}

\textit{Stationary state Entanglement Entropy ---} 
Finally, we compute analytically the entanglement entropy of the stationary state in the thermodynamic limit and for a large interval $\ell\gg 1$. The key insight is that for our non-Hermitian spin chain, much like for critical ground-state systems~\cite{peschel2004on,yates2018central,ares2015entanglement,ares2019sublogarithmic,fraenkel2021entanglement}, the leading behavior of the entanglement is controlled by the presence of singularities in $\Pi^\infty(k) \equiv \vec{\Pi}^\infty(k)\cdot \vec{\sigma}$ (cfr. Eq.~\eqref{eq:toepliz}). For $\gamma<\gamma_c$, $\Pi_x^\infty(k)$ exhibits sharp jumps at $k\simeq \pm k^*$ (See Fig.~\ref{fig:discontinuit}), which resemble those present for the ground state of critical fermionic systems, and can be interpreted as non-Hermitian Fermi points. On the other hand for $\gamma\ge \gamma_c$  all the components of the correlation matrix $\Pi^\infty(k)$ are smooth functions of momentum. As we show below, this different behavior is responsible for the behavior of the entanglement entropy as $\gamma$ is tuned through the critical point $\gamma_c$.

\begin{figure}[t]
	\includegraphics[width=\columnwidth]{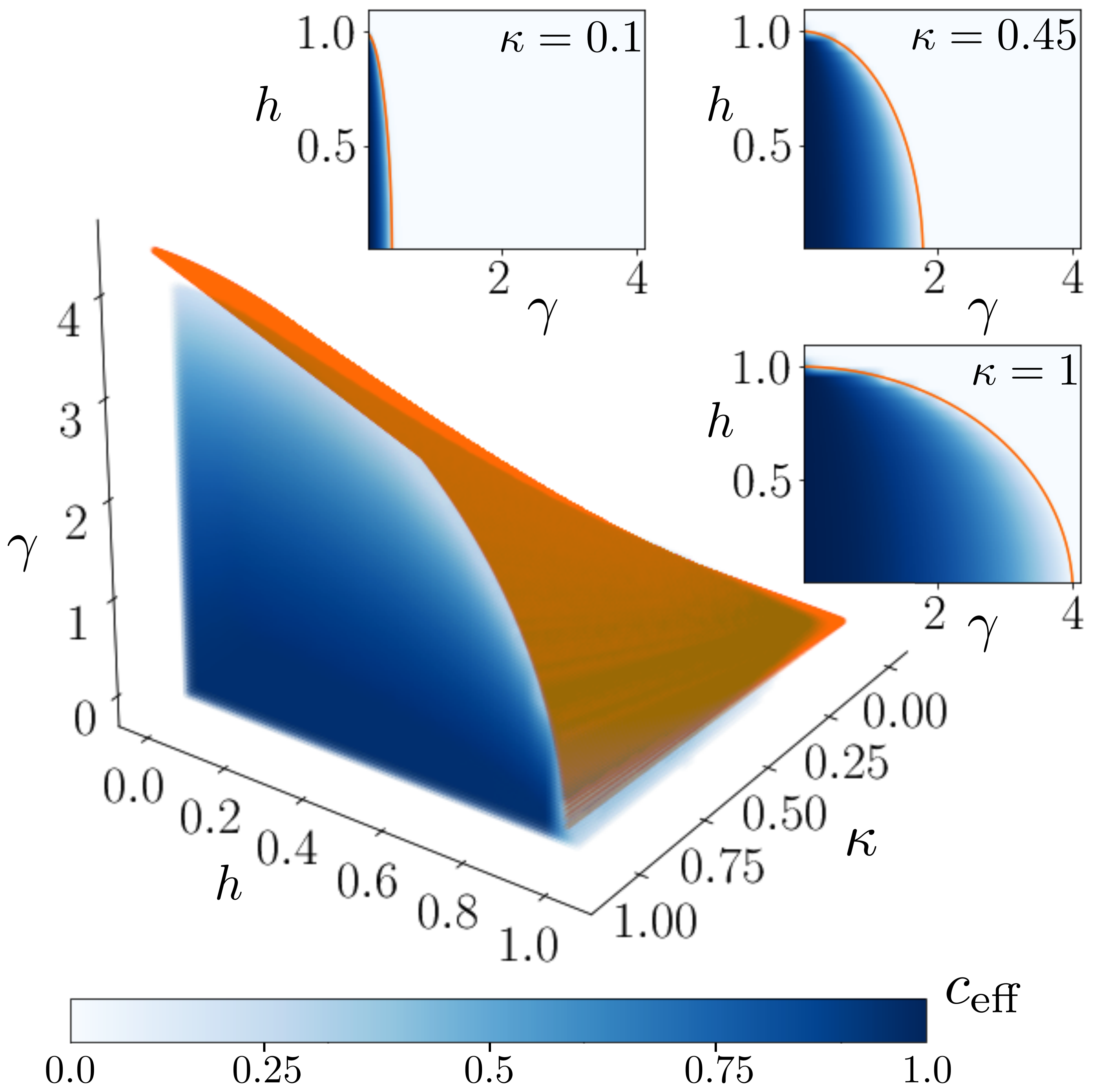}
	\caption{\label{fig:c_eff} Effective central charge $c_\mathrm{eff}$ varying the parameters of the non-Hermitian Hamiltonian. The boundary between the phase with logarithmic scaling of the entanglement entropy, leading to a finite $c_\mathrm{eff}$ and the area-law phase, corresponding to $c_\mathrm{eff}=0$ is given by the orange manifold.  }
\end{figure}
To proceed with focus on the gapless phase, $\gamma<\gamma_c$, and expand $\Pi^\infty(k)\simeq \tilde{\Pi}(k)$ around $k^*$ as 
\begin{equation}
 \tilde{\Pi}(k) \equiv \begin{pmatrix}
    -\beta  & \sqrt{1-\beta^2} \frac{k-k^*}{|k-k^*|}\\
    \sqrt{1-\beta^2} \frac{k-k^*}{|k-k^*|}  & \beta
    \end{pmatrix}\, . \label{eq:fermipoints}
\end{equation}
where we have neglected the subleading orders in $(k-k^*)$ -- which are responsible for subleading corrections, and introduced the parameter $\beta=\gamma/(4\kappa\sqrt{1-h^2})$.

After integrating over the momenta, we get
\begin{equation}
    \Gamma_{ij}^\ell \simeq -\beta \delta_{ij}\sigma^z + \sqrt{1-\beta^2} \left(\delta_{ij} - i(1-\delta_{ij})\frac{e^{-i k^*(i-j)}}{\pi (i-j)}\right)\sigma^x\;,\nonumber
\end{equation}
which can be diagonalised analytically to obtain the leading contribution to the entanglement entropy as detailed in the Supplemental Material~\cite{SM}. 
The final result is given by 
\begin{align}\label{eq:s_ell}
    S_\infty(\ell) &= (c_\mathrm{eff}/3) \ln \ell+\mathcal{O}(1)\\    
    c_\mathrm{eff} & = \frac{12}{\pi^2}\mathrm{Re}\int_0^1 d\lambda \Upsilon(\lambda) \frac{\lambda}{1-\lambda^2}\frac{\sqrt{1-\beta^2}}{\sqrt{\lambda^2-\beta^2}}\label{eq:ceff1}
\end{align}
where we have introduced the function $ \Upsilon(x) = -\frac{1+x}{2}\ln\left(\frac{1+x}{2}\right)- \frac{1-x}{2}\ln\left(\frac{1-x}{2}\right)$. 

Eq.~(\ref{eq:s_ell}-\ref{eq:ceff1}) are among the main results of this work, demonstrating that the stationary state entanglement entropy of the non-Hermitian anisotropic XY chain scales logarithmically with the size of the cut.  This logarithmic scaling, anticipated by the numerical data in Fig.~(\ref{fig:czz}), is shown here to emerge analytically from the level spacing of the spectrum of the matrix $\Gamma_{ij}^\ell$~\cite{SM}, consequence in the present setting of the parity of the system.
The prefactor $c_\mathrm{eff} $ depends non trivially from the parameters of the problem and it is reminiscent of the effective central charge of non-unitary conformal field theories~\cite{bianchini2015entanglement,couvreur2017entanglement,chang2020entanglement}. In particular  it vanishes continuously for $\beta\to 1$
with a derivative diverging as $\partial_\beta c_\mathrm{eff}\simeq (1-\beta^2)^{-1/2}$. This corresponds to manifold in the parameter space $(\gamma,h,\kappa)$ leading to the phase diagram shown in Fig.~\ref{fig:c_eff}. For $\gamma<\gamma_c(\kappa,h)$ the entanglement scales logarithmically while for $\gamma\ge \gamma_c$ the effective central charge vanishes and the entanglement displays area law. Our analytical results therefore demonstrate an entanglement transition in the non-Hermitian anisotropic XY chain.
Furthermore, we note that the same diverging behavior is present for the monitored quantum system and the non-Hermitian quasi-particle phenomenology presented in Ref.~\cite{turkeshi2021entanglement}. 

The exact analytical result for the effective central charge allows to discuss and clarify the limits $\gamma\to 0$ and $\kappa\to 0$, corresponding respectively to a vanishing dissipation and vanishing anisotropy (leading to the XX model). In the former case we have $\lim_{\gamma\to 0} c_\mathrm{eff}=1$, a value that has to be still interpreted as effective central charge and not to be confused with the result obtained for the critical ground-state anisotropic XY model ($c_\mathrm{Ising}=1/2$) which cannot be obtained within our nonequilibrium framework. Starting with $\gamma=0$ to  begin with would in fact lead to a linear scaling in time of the entanglement. Furthermore, the limit $\kappa\to 0$ (the isotropic XX chain) suggests the model is always in an area law. This may have important consequences for the measurement induced transitions on the associated (non-postselected) dynamics. 

\textit{Conclusion ---} In this Letter we have studied the entanglement and correlation spreading in a non-Hermitian anisotropic XY spin chain, after sudden switching of the dissipation (or measurement) rate. Using free fermion techniques we have solved exactly the model and computed the dynamics of the entanglement entropy and of the spin-spin correlation function in the thermodynamic limit.  The model shows a transition in the spectrum of decay modes, going from gapless around two k points for $\gamma<\gamma_c$ to gapped throughout the Brillouin zone for $\gamma>\gamma_c$. We have demonstrated that this transition drives a sharp change in the dynamical properties of the model, which for dissipation $\gamma<\gamma_c$ displays logarithmic entanglement growth and correlation spreading violating the conventional Lieb-Robinson bound while for $\gamma>\gamma_c$ behaves as a conventional off critical quantum spin chain with finite-correlation length. Furthermore we have computed analytically the leading contribution to the stationary state entanglement entropy in the gapless phase, demonstrating its logarithmic scaling with the size of the cut and providing a closed-form expression for the effective central charge. We have shown the latter to vanish continuously at the spectral transition $\gamma_c$ which therefore identifies also a sharp entanglement transition from logarithmic to area law scaling. Our results have profound implications both for non-Hermitian quantum many-body systems, which are shown to possess non-trivial entanglement dynamics whose general pattern and classification calls for further studies, and for measurement-induced entanglement transitions under post-selected dynamics~\cite{potter2021entanglement,lunt2021quantum}, and space-time dual dynamics~\cite{ippoliti2021postselectionfree,lu2021spacetime}, for which our model provides an exactly solvable case.

\begin{acknowledgements}
This work was supported by the ANR grant ``NonEQuMat''(ANR-19-CE47-0001). We acknowledge computational resources on the Coll\'ege de France IPH cluster. 
\end{acknowledgements}


\bibliography{new_nhtqim}

\bibliographystyle{apsrev4-2} 
\normalem

\widetext
\clearpage
\begin{center}
\textbf{\large \centering Supplemental Material:\\ Entanglement and Correlation Spreading in non-Hermitian Spin Chains}
\end{center}

\setcounter{equation}{0}
\setcounter{figure}{0}
\setcounter{table}{0}
\setcounter{page}{1}
\renewcommand{\theequation}{S\arabic{equation}}
\setcounter{figure}{0}
\renewcommand{\thefigure}{S\arabic{figure}}
\renewcommand{\thepage}{S\arabic{page}}
\renewcommand{\thesection}{S\arabic{section}}
\renewcommand{\thetable}{S\arabic{table}}
\makeatletter

\renewcommand{\thesection}{\arabic{section}}
\renewcommand{\thesubsection}{\thesection.\arabic{subsection}}
\renewcommand{\thesubsubsection}{\thesubsection.\arabic{subsubsection}}

In this Supplemental Material we discuss:
\begin{enumerate}
    \item[S1] How the non-Hermitian quantum quench of interest is achieved from a quantum jump stochastic Schr\"odinger equation;
    \item[S2] How the exact solution of the wave-function dynamics is obtained by means of fermionization;
    \item[S3] The computation of the dynamics and of the exact stationary state correlation functions;
    \item[S4] The computation of the dynamics and of the exact stationary state entanglement entropy.
\end{enumerate}

\section{S1 Non-Hermitian quantum quench from quantum jumps}
\label{sec:intro}
We consider the XY spin chain coupled to a measuring apparatus which stochastically detect quantum jump events. The XY Hamiltonian ${\hat H=\hat H(\kappa,h)}$ is given by
\begin{align}
	\hat{H}(\kappa,h) = -\sum_{i=1}^L \left(\frac{1+\kappa}{2}\hat\sigma^x_i \hat\sigma^x_{i+1}+ \frac{1-\kappa}{2}\hat\sigma^y_i \hat\sigma^y_{i+1} + h \hat\sigma^z_i\right),
\end{align}
where $h$ is the transverse magnetic field, $\kappa$ is the anisotropy parameter, and periodic boundary conditions (\textbf{PBC}) are inferred. The apparatus measures $\hat{n}_i\equiv (1+\hat{\sigma}^z_i)/2$ with rate $\gamma$, such that the evolution of the system follows the quantum jump (\textbf{QJ}) equation
\begin{align}
	d|\psi_t\rangle = -i H dt|\psi_t\rangle - \frac{\gamma}{2}dt \sum_{i=1}^L (\hat{n}_i-\langle \hat{n}_i\rangle_t ) |\psi_t\rangle + \sum_{i=1}^L d\mathcal{N}^i_t \left(\frac{\hat{n}_i}{\sqrt{\langle \hat{n}_i\rangle_t}} - 1\right) |\psi_t\rangle.
\end{align}
The jumps are a Poisson stochastic process $d\mathcal{N}^i_t = 0,1$, with $d\mathcal{N}^i_t d\mathcal{N}^j_t = \delta^{ij} d\mathcal{N}^i_t$ and $\overline{d\mathcal{N}^i_t} = \gamma \langle \hat{n}_i\rangle_t dt$. Each history $\mathcal{N}_t$ fix a unique quantum trajectory.
If we post-select the one with no jump occurring (no-click limit), the dynamics is fixed by the non-Hermitian Hamiltonian
\begin{align}
	\hat{H}_\mathrm{eff} = \hat{H}_\mathrm{eff}(\kappa,h,\gamma) = \hat{H}(\kappa,h) -i\frac{\gamma}{4}\sum_{m=1}^L \hat{\sigma}_z.\label{eq:nonhermham}
\end{align}

We start from an initial state which is the ground state $|\Psi_0\rangle$ of the XY Hamiltonian $\hat{H}(\kappa_0,h_0)$, and consider a quench where the new Hamiltonian which governs the time evolution is $\hat{H}(\kappa,h)$ and the measurement apparatus is coupled to the system. After the aforementioned post-selection, the resulting dynamics is fixed by the quench $\hat{H}(\kappa_0,h_0)\mapsto \hat{H}_\mathrm{eff}(\kappa,h,\gamma)$, which is the object of interest for this study.

The formal solution of the post-selected equation of motion
\begin{align}
	\frac{d}{dt}|\Psi(t)\rangle = -i \hat{H}_\mathrm{eff} |\Psi(t)\rangle -\frac{i}{2}\langle \hat{H}^\dagger_\mathrm{eff} - H_\mathrm{eff}\rangle_t|\Psi(t)\rangle,\label{eq:nonhermevo}
\end{align}
is formally given by
\begin{align}
	|\Psi(t)\rangle = \frac{e^{-i \hat{H}_\mathrm{eff} t}|\Psi_0\rangle}{||e^{-i \hat{H}_\mathrm{eff} t}|\Psi_0\rangle||}.\label{eq:abstractsol}
\end{align}
In the following we solve exactly the dynamics of this model by means of free fermion techniques and using the quadratic nature of the problem. 
We conclude this section by noting that the system is closed but not isolated. In the spirit of Ref.~\cite{gisin1984quantum}, given the decomposition $\hat{H}_\mathrm{eff} = \hat{H}- i\gamma \hat{\Gamma}$ and denoting $\langle \circ \rangle_t \equiv \langle \Psi(t)|\circ|\Psi(t)\rangle$ the expectation value over the state, we have
\begin{align}
	\frac{d}{dt}\langle \hat{H}\rangle_t = - \gamma \left\langle \Psi(t)\left| \left\lbrace \hat{H}-\langle \hat{H}\rangle_t,\hat\Gamma-\langle \hat\Gamma\rangle_t\right\rbrace\right|\Psi(t)\right\rangle\le 0,\label{eq:dissipation}
\end{align}
and the equality holds only for $\gamma=0$. For this reason, the measurement rate $\gamma$ is interpreted as a dissipation rate in the Main Text.

\section{S2 Solution of the dynamics of the wave-function}
\label{sec:solwave}
In this section we exactly solve the dynamics of the many-body wave-function. First we rephrase the model in fermionic language through the Jordan-Wigner transformation. Afterward, we diagonalize the model in momentum-space, and lastly we express the time-dependent state in terms of exactly computable factors.

\subsection{Fermionization through Jordan-Wigner transformation}
To solve the dynamics for translational invariant systems, we first map the problem through Jordan-Wigner transformation, which is given by
\begin{equation}
	\begin{cases}
		\hat\sigma^x_m \mapsto e^{i\pi \sum_{r=1}^{m-1}\hat{c}^\dagger_r\hat{c}_r}(\hat{c}^\dagger_m + \hat{c}_m)\\
		\hat\sigma^y_m \mapsto ie^{i\pi \sum_{r=1}^{m-1}\hat{c}^\dagger_r\hat{c}_r}(\hat{c}^\dagger_m - \hat{c}_m)\\
		\hat\sigma^z_m \mapsto 1-2 \hat{n}_m,\quad \hat{n}_m \equiv \hat{c}^\dagger_{m}\hat{c}_{m}
	\end{cases}
	\label{eq:JW}
\end{equation}
with $\hat{c}_m$ ($\hat{c}^\dagger_m)$) fermionic annihilation (creation) operators, satisfying  $\{\hat{c}_m,\hat{c}^\dagger_r\}=\delta_{m,r}$. The transformed non-Hermitian Hamiltonian Eq.~\eqref{eq:nonhermham} is 
\begin{align}
	\hat{H}_\mathrm{eff} &= -\sum_{m=1}^{L-1} (\hat{c}^\dagger_{m}\hat{c}_{m+1} + \kappa \hat{c}^\dagger_{m}\hat{c}^\dagger_{m+1} + \mathrm{h.c.}) - \left(2 h+i \frac{\gamma}{2}\right) \sum_{m=1}^L \hat{c}^\dagger_{m}\hat{c}_{m} + 2L h \nonumber \\
	& \qquad +  (-1)^{\hat{N}}(\hat{c}^\dagger_{L}\hat{c}_{1} + \kappa\hat{c}^\dagger_L \hat{c}^\dagger_1 + \mathrm{h.c.}).
\end{align}
The boundary term depends on the fermionic parity $(-1)^{\hat{N}}$ (here $\hat{N}= \sum_{m=1}^L \hat{n}_m$ is the number of fermions), and arises from the explicit breaking of the periodic boundary condition of the spin chain, as Eq.~\eqref{eq:JW} fixes a starting point of the fermionic chain. 

If the number of fermions is even, then anti-periodic boundary conditions (\textbf{ABC}) apply; if instead the number of fermions is odd, PBCs apply. 
In particular, introducing the projectors onto the fermionic parity sectors
\begin{equation}
	\hat{P}_\pm = \frac{1}{2}\left(1\pm e^{i\pi \hat{N}}\right),
\end{equation}
we have 
\begin{align}
	H_\mathrm{eff} = P_+ H_\mathrm{eff} P_+ + P_- H_\mathrm{eff} P_- = H^+_\mathrm{eff} + H^-_\mathrm{eff},
\end{align}
where $H^\pm_\mathrm{eff} = P_\pm H_\mathrm{eff} P_\pm$ are the restriction of the spin Hamiltonian to the sectors with parity $\pm 1$ respectively. 
Throughout this work we are interested in the quench dynamics starting from a state with well defined fermionic parity. In this case, only one of the Hamiltonians $\hat{H}^\pm_\mathrm{eff}$ acts non-trivially on the system. For convenience we shall consider an even-parity state, and use the slight abuse of notation $\hat{H}_\mathrm{eff}\equiv \hat{H}^+_\mathrm{eff}$.

\subsection{Diagonalization and dynamics through Fourier-transform}
The Fourier transform is given by
\begin{align}
	\begin{cases}
		\displaystyle\hat{c}_m = \frac{e^{\frac{i \pi}{4}}}{\sqrt{L}} \sum_k e^{ikm} \hat{c}_k,\\
		\displaystyle\hat{c}_k = \frac{e^{\frac{-i \pi}{4}}}{\sqrt{L}} \sum_{m=1}^L e^{-ikm} \hat{c}_m,
	\end{cases} 
\end{align}
where the overall phase $\exp(\pm i\pi/4)$ does not affect the canonical anti-commutation relations, and it is introduced for latter simplifications. The choice of momenta $k$ depends on the parity sector. If PBC hold for the fermionic chain, the allowed momenta are the roots of $\exp(ikL)=1$, that is
\begin{align}
	\text{OBC}: \qquad k\in \mathcal{K}_+ =\left\lbrace \frac{2 n \pi}{L} \Big|n = -\frac{L}{2}+1,\dots,0,\dots,\frac{L}{2}\right\rbrace,
\end{align}
whereas if ABC hold, the allowed momenta are given by $\exp(ikL) = -1$, hence 
\begin{align}
	\text{ABC}: \qquad k\in \mathcal{K}_-=\left\lbrace \pm\frac{(2 n-1) \pi}{L} \Big| n = 1,\dots,\frac{L}{2}\right\rbrace.
\end{align}
For the dynamics of interest, we consider a state with even number of fermions, and hence ABCs. 
Then, the resulting fermionic Hamiltonian is Fourier transformed to the following 
\begin{align}
	\hat{H}_\mathrm{eff} &= -\sum_{k\in\mathcal{K}_-} (i e^{ik}\kappa \hat{c}^\dagger_k  \hat{c}^\dagger_{-k} + \hat{c}^\dagger_k\hat{c}_k e^{ik} + \mathrm{h.c.}) -\sum_{k\in \mathcal{K}_-}\left[\left(2h+i\frac{\gamma}{2}\right) \hat{c}^\dagger_k \hat{c}_k-2h\right].\label{eq:ftnherm}
\end{align}
where $\hat{c}_k$ ($\hat{c}^\dagger_k$) are the annihilation (creation) fermionic operators with momentum $k$. We define the positive momenta
\begin{align}
	\mathcal{K} \equiv \{ k\in \mathcal{K}_-| k>0\}.
\end{align}
Then, using the parity of trigonometric functions we have
\begin{align}
	\hat{H}_\mathrm{eff} = \sum_{k\in\mathcal{K}} \left[\begin{pmatrix}\hat{c}_k^\dagger&\hat{c}_{-k}\end{pmatrix} M_k \begin{pmatrix}\hat{c} _k&\hat{c}^\dagger_{-k}\end{pmatrix}^T - i\frac{\gamma}{2} - 2 \cos(k)\right],
\end{align}
with the matrix 
\begin{align}
	M_k &= -\varepsilon_k \sigma^z +\Delta_x \sigma^x = \begin{pmatrix}
		-\varepsilon_k & \Delta_k\\
		\Delta & +\varepsilon_k
	\end{pmatrix},\nonumber\\ \varepsilon_k &= 2\cos k + 2h + i\frac{\gamma}{2},\qquad \Delta_k = 2 \kappa \sin k.
\end{align}
The constant term $\Lambda_\mathrm{shift}\equiv -i\gamma/2 - 2\cos k$ can be eliminated by noting: (i) $2\cos k$ is responsible for a global phase of the wave-function, (ii) the imaginary constant $\gamma/2$ is simplified by the same constant term arising in the feedback in Eq.~\eqref{eq:nonhermevo} (equivalently, see Eq.~\eqref{eq:abstractsol}). 
Hence, with a slight abuse of notation, we neglect them from now on.

The final expression for the non-Hermitian Hamiltonian is
\begin{align}
	\hat{H}_\mathrm{eff} = \sum_{k\in \mathcal{K}} \hat{H}_k,\qquad\text{with}\qquad
	\hat{H}_k = \begin{pmatrix}\hat{c}_k^\dagger&\hat{c}_{-k}\end{pmatrix} M_k \begin{pmatrix}\hat{c}_k\\\hat{c}^\dagger_{-k}\end{pmatrix}.\label{eq:momkham}
\end{align}
The $k$-momentum non-Hermitian Hamiltonian $\hat{H}_k$ acts on the space generated by 
\begin{equation}
	|0\rangle, \hat{c}^\dagger_k|0\rangle, \hat{c}^\dagger_{-k}|0\rangle, \hat{c}^\dagger_k \hat{c}^\dagger_{-k}|0\rangle,
\end{equation}
but, due to the conservation of the fermionic parity, only its action on the manifold $\{|0\rangle,\hat{c}^\dagger_k \hat{c}^\dagger_{-k}|0\rangle\}$ is non-trivial. Within this subspace, upon relabelling of the basis, the action is equivalent to that of the matrix $M_k$.
The spectrum of $M_k$ is given by $\Lambda_k = \pm \sqrt{\varepsilon_k^2+ \Delta_k^2} \equiv \pm E_k  \pm i\Gamma_k$. 
The convention on the sign of $\Lambda_k$ can be independently chosen for every momentum $k$~\cite{lee2014heralded}, and in particular can be fixed in such a way that $\Gamma_k\le 0$.
Introducing for convenience $\tilde\gamma=\gamma/4$, a simple algebraic computation gives
\begin{align}
	\Xi_k &= h^2+\kappa^2-\tilde{\gamma}^2 + (1-\kappa^2)\cos^2(k) + 2h\cos(k)\\
	E_k &= \sqrt{2} \sqrt{ \Xi_k+ \sqrt{\Xi_k^2 + 4 \tilde{\gamma}^2 (\cos(k)+h)^2}}\\
	\Gamma_k &= 4 \frac{\tilde\gamma (\cos(k)+h)}{E_k}.
\end{align}
In particular, we identify modes with gapless imaginary part for $k=\pm\arccos(-h)$.

We are now in position to compute the exact dynamics of the many-body quantum state. Since the state $|\Psi_0\rangle$ is translational invariant and with a well defined parity, then its dynamics can be decomposed in $N/2$ separate ones 
\begin{align}
	|\Psi_0\rangle &= \prod_{k\in \mathcal{K}}|\psi_k(0)\rangle,\qquad |\Psi_t\rangle = \prod_{k\in \mathcal{K}}|\psi_k(t)\rangle\nonumber \\
	|\psi_k(t)\rangle &= \frac{|\tilde{\psi}_k(t)\rangle} {|||\tilde{\psi}_k(t)\rangle||},\qquad i\frac{d}{dt}|\tilde{\psi}_k(t)\rangle = \hat{H}_k |\tilde{\psi}_k\rangle,\label{eq:exact}
\end{align}
where the $k$-momentum states are expressed as $|\tilde{\psi}_k(t)\rangle = u_k(t)|0\rangle + v_k(t) \hat{c}^\dagger_k \hat{c}^\dagger_{-k}|0\rangle$.
In matrix notation we have
\begin{align}
	i\frac{d}{dt}\begin{pmatrix}
		u_k(t)\\v_k(t)
	\end{pmatrix} = M_k \begin{pmatrix}
		u_k(t)\\v_k(t)
	\end{pmatrix}
\end{align}
which is a linear ordinary differential equation and can be solved exactly through standard methods.
Introducing $u_k^0\equiv \cos(\theta_k^0) \equiv u_k(0) $ and $v_k^0\equiv\sin(\theta_k^0) \equiv v_k(0)$ (this decomposition is always possible as the initial state is normalized and the initial Hamiltonian is real and symmetric) we have
\begin{align}
	u_k(t) &= \cos(\theta_k^0)\cos(\Lambda_k t) -i \cos(\Phi_k)\sin(\Lambda_k t)\nonumber \\
	v_k(t) &= \sin(\theta_k^0)\cos(\Lambda_k t) +i \sin(\Phi_k)\sin(\Lambda_k t)\label{eq:wvsolk}
\end{align}
where $\theta_k = \arccos (E_k/\Lambda_k)$ and $\Phi_k\equiv \theta_k - \theta_k^0$.

Eq.~\eqref{eq:wvsolk} is the exact solution of a single momentum sector $k$. Plugging it back in Eq.~\eqref{eq:exact} the final solution is given by the BCS wave-function
\begin{align}
	|\Psi_t\rangle = \prod_{k\in \mathcal{K}} \left(\frac{u_k(t) + v_k(t) \hat{c}^\dagger_k \hat{c}^\dagger_{-k}}{\sqrt{|u_k(t)|^2+|v_k(t)|^2}} \right)|0\rangle. \label{eq:exactsolution}
\end{align}
We conclude with few remarks. First, the dynamics preserve the Gaussianity as expected by the quadratic form of the non-Hermitian Hamiltonian. The environment feedback acts as a renormalization which ensures the state is always pure. 
Factoring out the term with the slowest decay modes, it is clear the stationary state is the eigenstate of $\hat{H}_\mathrm{eff}$ with the smallest (in absolute value) imaginary part. Hence, the presence/absence of the (imaginary part) gapless modes is crucial in determining the dynamical and stationary state properties of the system. 
Secondly, an exact solution can be investigated also for the open boundary condition case, using the techniques in Ref.~\cite{turkeshi2021diffusion}. For the sake of conciseness, we do not present this solution in this paper, the main difference being the presence of two degenerate stationary states with the same $\Gamma_k$, but different $E_k$, resulting in oscillatory behavior in the stationary state (See the numerics in Ref.~\cite{turkeshi2021measurementinduced} for the $h=0$, $\kappa=1$ case).
These states correspond, in the translational invariant case, to those with the slowest decay modes of the even/odd sectors, which are not anymore degenerate due to the boundary interaction.

\section{S3 Dynamics of the correlation functions}
The Gaussian form of the state Eq.~\eqref{eq:exactsolution} allows us to introduce the Nambu modes $\hat{\eta}_k(t) = u_k(t) c_k - v_k(t) c_{-k}^\dagger$ such that $\hat{\eta}_k|\Psi_t\rangle = 0$ for any $t,k$. 
(An aside here is that, $u_k(t)$ and $v_k(t)$ in Eq.~\eqref{eq:exactsolution} can be always chosen to generate complex Bogoliubov rotations $u_k(t)^2+v_k(t)^2=1$, as the non-Hermitian Hamiltonian is complex symmetric $\hat{H}_\mathrm{eff} = \hat{H}_\mathrm{eff}^T$). 
Inverting this transformation we have the following
\begin{equation}
    \label{eq:complexbogo}
    \hat{c}_k = \frac{\overline{u}_k(t) \hat{\eta}_k + v_k(t) \hat{\eta}_{-k}^\dagger}{|u_k(t)|^2 + |v_k(t)|^2},
\end{equation}
where we use the overline $\overline{\circ}$ to denote the complex conjugate of a number. 
An important caveat is that the opeartors $\hat{\eta}_k$ and $\hat{\eta}_{k}^\dagger$ are not canonical conjugate, as $\{\hat{\eta}_k,\hat{\eta}_{k'}^\dagger\} = \delta_{k,k'} (|u_k(t)|^2 + |v_k(t)|^2)$, which is a consequence of the non-Hermitian nature of the problem.

By virtue of Guassianity and employing Wick's theorem, the correlation functions and the entanglement properties are fully encoded in the two-point function of the Majorana fermions, which up to a phase are given by the operators  $\hat{A}=\hat{c}_n^\dagger + \hat{c}_n$ and $\hat{B}_n = \hat{c}_n^\dagger - \hat{c}_n$~\cite{peschel2004on,jin2005entanglement}.

A simple computation gives the following
\begin{align}
	 \langle \Psi_t| \hat{A}_m \hat{A}_n| \Psi_t\rangle &= \delta_{mn} + \frac{4i}{N}\sum_{k\in \mathcal{K}} \sin(k(n-m))\left(\frac{\mathrm{Im}(u_k(t)\overline{v}_k(t))}{|u_k(t)|^2+ |v_k(t)|^2} \right)\\
	 \langle \Psi_t| \hat{B}_m \hat{B}_n| \Psi_t\rangle &= -\delta_{mn} + \frac{4i}{N}\sum_{k\in \mathcal{K}} \sin(k(n-m))\left(\frac{\mathrm{Im}(u_k(t)\overline{v}_k(t))}{|u_k(t)|^2+ |v_k(t)|^2} \right)\\
	 \langle \Psi_t| \hat{B}_m \hat{A}_n| \Psi_t\rangle &= -  \langle \Psi_t| \hat{A}_m \hat{B}_n| \Psi_t\rangle \nonumber \\
	 &= -\frac{2}{N}\sum_{k\in \mathcal{K}} \cos(k(n-m))\left(\frac{|u_k(t)|^2- |v_k(t)|^2}{|u_k(t)|^2+ |v_k(t)|^2} \right) \nonumber \\ &\qquad + \frac{4}{N}\sum_{k\in \mathcal{K}} \sin(k(n-m))\left(\frac{\mathrm{Re}(u_k(t)\overline{v}_k(t))}{|u_k(t)|^2+ |v_k(t)|^2} \right)\label{eq:correfun}
\end{align}
where the $\sin$ and $\cos$ factors are displayed to highlight the parity of these correlations. 

\begin{figure}[t]
    \centering
    \includegraphics[width=0.9\columnwidth]{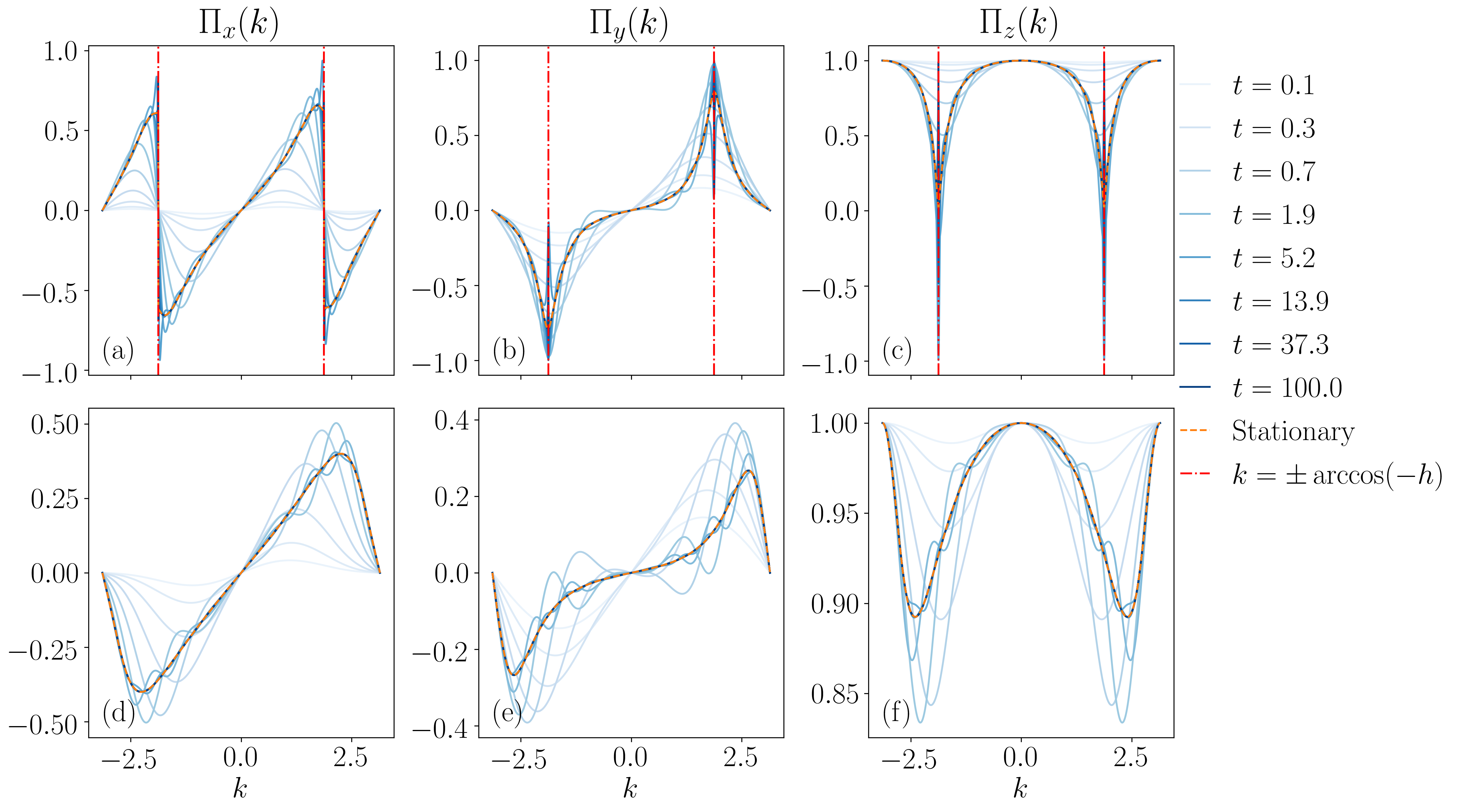}
    \caption{\label{fig:testcormatl} Simulation of the correlation matrix $\Pi(k)$ at different times for $\kappa=0.4$, $\gamma=1.2.$ and  $h=0.3$ (a-c), and for $\kappa=0.4$, $\gamma=1.2.$ and  $h=0.3$ (d-f).  A singular behavior is exhibited for $k=k^*=\arccos(-h)$ in the imaginary gapless phase: for the $x$-component, this is given by a sign singularity, whereas for the other two functions these are absolute value singularities. In the imaginary gapped phase, these singularities are smoothed away.}
\end{figure}

Taking the thermodynamic limit $N\to \infty$ the above sums are transformed into integrals, and we have
\begin{align}
    \langle \Psi_t| \hat{A}_m \hat{A}_n| \Psi_t\rangle &= \delta_{mn} - \frac{1}{2\pi }\int_{-\pi}^\pi \, dk e^{-i k (n-m)} \Pi_z(k), \\
	 \langle \Psi_t| \hat{B}_m \hat{B}_n| \Psi_t\rangle &= -\delta_{mn} -  \frac{1}{2\pi }\int_{-\pi}^\pi \, dk e^{-i k (n-m)} \Pi_z(k),\\
	 \langle \Psi_t| \hat{B}_m \hat{A}_n| \Psi_t\rangle &= -\frac{i}{2\pi }\int_{-\pi}^\pi \, dk e^{-i k (n-m)} (\Pi_x(k)+i \Pi_y(k)) , \\ \Pi_x(k) =  2\frac{\mathrm{Re}(u_k(t)\overline{v}_k(t))}{|u_k(t)|^2+ |v_k(t)|^2}, \qquad &\Pi_y(k) = -\frac{|u_k(t)|^2- |v_k(t)|^2}{|u_k(t)|^2+ |v_k(t)|^2}, \qquad  \Pi_z(k) = 2 \frac{\mathrm{Im}(u_k(t)\overline{v}_k(t))}{|u_k(t)|^2+ |v_k(t)|^2}. \label{eq:correfun2}
\end{align}
The function $\Pi_\alpha(k)$ are introduced for later convenience, and their momentum dependence determines how the spatial correlations behave at large distance.
Before proceeding, we also consider the stationary state, which as already stated, is the eigenstate of the non-Hermitian Hamiltonian with the lowest decay rate $\Gamma_k$. For this state we have~\cite{lee2014heralded}
\begin{equation}
    u_k(\infty) = \frac{- \varepsilon_k - \mathrm{sign}(\Gamma_k) \Lambda_k}{C},\qquad 
    v_k(\infty) = \frac{- \Delta_k}{C},
\end{equation}
where $C^2 = u^2_k(\infty)+v^2_k(\infty)$. We preliminary plot the time-evolution of the $\Pi_\alpha(k)$ functions and of their stationary state in Fig.~\ref{fig:testcormatl}. As we see, a separation is present depending on the gapless/gapped nature of $\Gamma_k$ with momentum: in the former (panels (a-c)) there is a late-time singular behavior in $\Pi_\alpha(k)$, whereas in the latter (panels (d-f)) $\Pi_\alpha(k)$ present smooth momentum behavior. 
We also note that at early time the singular behavior is not felt by the system, as the decay modes do not still act actively on the system. 
This explains the cross-over effects present at early times, or at late time for relatively small system sizes (\textit{e.g.}, in the numerics of Ref.~\cite{turkeshi2021measurementinduced,turkeshi2021entanglement}).
In fact, taking the thermodynamic limit is relevant for a detailed description of the phase characterizing the system. At the level of correlation functions, spurious effects are present for the long time dynamics if the momentum separation is not finely resolved.

In the Main Text we discuss the evolution of the connected spin-spin correlation function 
\begin{equation}
    C^\mathrm{zz}_x(t) = \langle \Psi_t| \hat\sigma^z_0\hat\sigma^z_x |\Psi_t\rangle -  \langle \Psi_t| \hat\sigma^z_0|\Psi_t\rangle\langle \Psi_t|\hat\sigma^z_x |\Psi_t\rangle = -  \langle \Psi_t| \hat{A}_0 \hat{B}_x |\Psi_t\rangle \langle \Psi_t| \hat{A}_x \hat{B}_0 |\Psi_t\rangle -  \langle \Psi_t| \hat{A}_0 \hat{A}_x |\Psi_t\rangle \langle \Psi_t| \hat{B}_0 \hat{B}_x |\Psi_t\rangle,
\end{equation}
where the last expression is obtained after simple manipulations with the Jordan-Wigner transformation and using the Wick's theorem~\cite{lee2014heralded}.

\subsection{Stationary state}
Computing a closed form for the time evolving observables in Eq.~\eqref{eq:correfun2} is in general a hard task, due to the non-trivial and non-universal dependence on the initial state. A simplified setup is the computation of the stationary state correlations which can be achieved by working out a closed form for 
\begin{align}
    I^\alpha_l(\infty) = \frac{1}{2\pi}\int_{-\pi}^{\pi}\, dk e^{-i k l } \Pi_\alpha(k;t=\infty),\qquad \alpha=x,y,z.\label{eq:integrals}
\end{align}
As shown in Fig.~\ref{fig:testcormatl} the integrand manifest a smooth part, and depending on the phase, a singular behavior around $k\simeq k^* = \pm\arccos(-h) $. 
For the gapless phase, the singularity at $k\simeq k^*$ is responsible for algebraic decay of spatial correlations, whereas for the gapped phase the $I^\alpha_l(\infty)$ are exponentially decaying. In the following we detail the computation for the gapless phase, which we obtain at leading order in $(k-k^*)$.

Hence it is convenient to first expand the various functions involved in the computation around this transition point. Up to corrections $\mathcal{O}((k-k^*)^2)$, we have
\begin{align}
    \Delta_k & \simeq 2\kappa \left(\sin k^* + \cos k^* (k-k^*) \right) = 2 \kappa \left(\sqrt{1-h^2} -h (k-k^*)\right),\\
    \varepsilon_k & \simeq 2 (\cos k^* - \sin k^* (k-k^*) + h + i \tilde\gamma) = 2 \left(i\tilde\gamma - (k-k^*) \sqrt{1-h^2}\right),\\
    \Lambda_k & \simeq 2\sqrt{ - \tilde\gamma^2 + \kappa^2(1-h^2) - 2 i \tilde\gamma (k-k^*) \sqrt{1-h^2} - 2 h\sqrt{1-h^2}\kappa^2 (k-k^*)}\nonumber\\
    &= 2 \sqrt{\kappa^2(1-h^2)-\tilde\gamma^2}-2 (k-k^*)\frac{i\tilde\gamma \sqrt{1-h^2} + \sqrt{1-h^2}\kappa^2}{\sqrt{\kappa^2(1-h^2)-\tilde\gamma^2}},\\
    E_k &\simeq 2 \sqrt{\kappa^2(1-h^2)-\tilde\gamma^2} - 2(k-k^*) \frac{\kappa^2 h \sqrt{1-h^2}}{\sqrt{\kappa^2(1-h^2)-\tilde\gamma^2}},\\
    \Gamma_k &\simeq  -2 (k-k^*)\frac{\tilde\gamma \sqrt{1-h^2}}{\sqrt{\kappa^2(1-h^2)-\tilde\gamma^2}}.
\end{align}

\begin{figure}[t]
    \centering
    \includegraphics[width=.8\columnwidth]{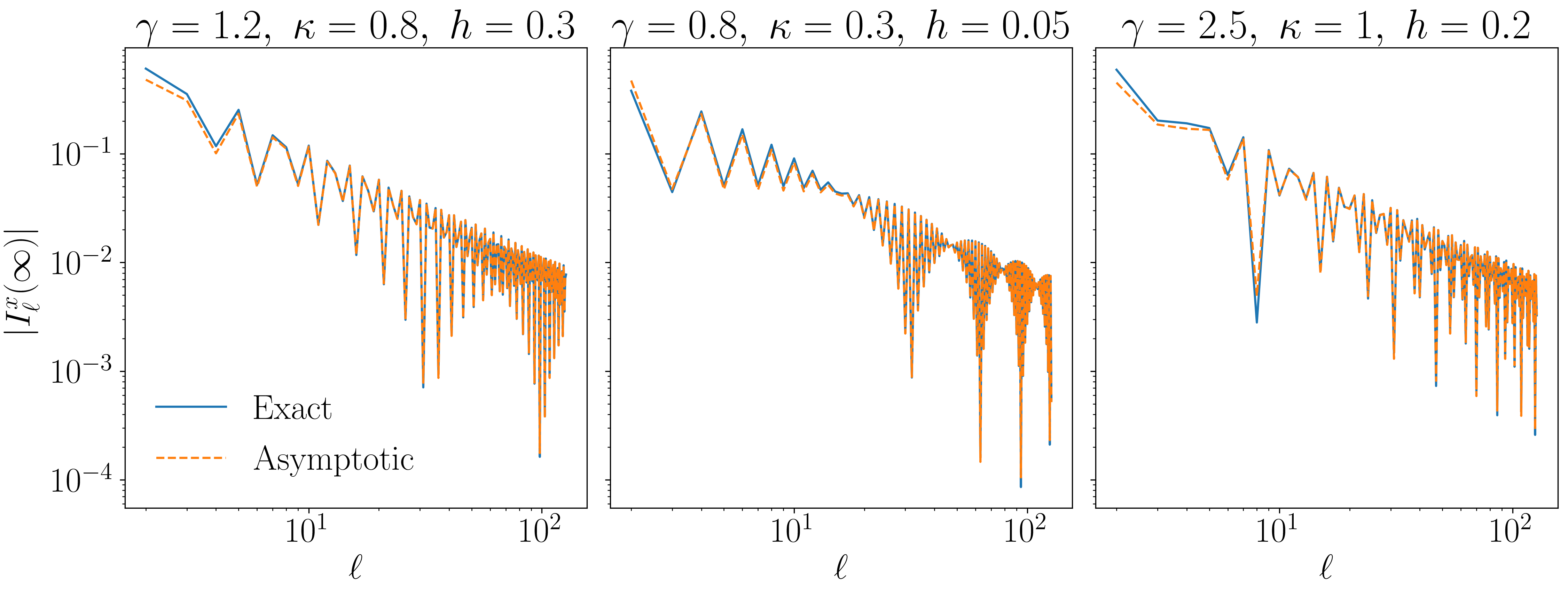}
    \caption{\label{fig:Ix} Comparison between the exact expression for $I^x_l(\infty)$ and the asymptotic behaviour Eq.~\eqref{eq:IXEXACT} within the imaginary gapless phase. The agreement is quantitative already for modest distance $l$. }
\end{figure}

\begin{figure}[t]
    \centering
    \includegraphics[width=.8\columnwidth]{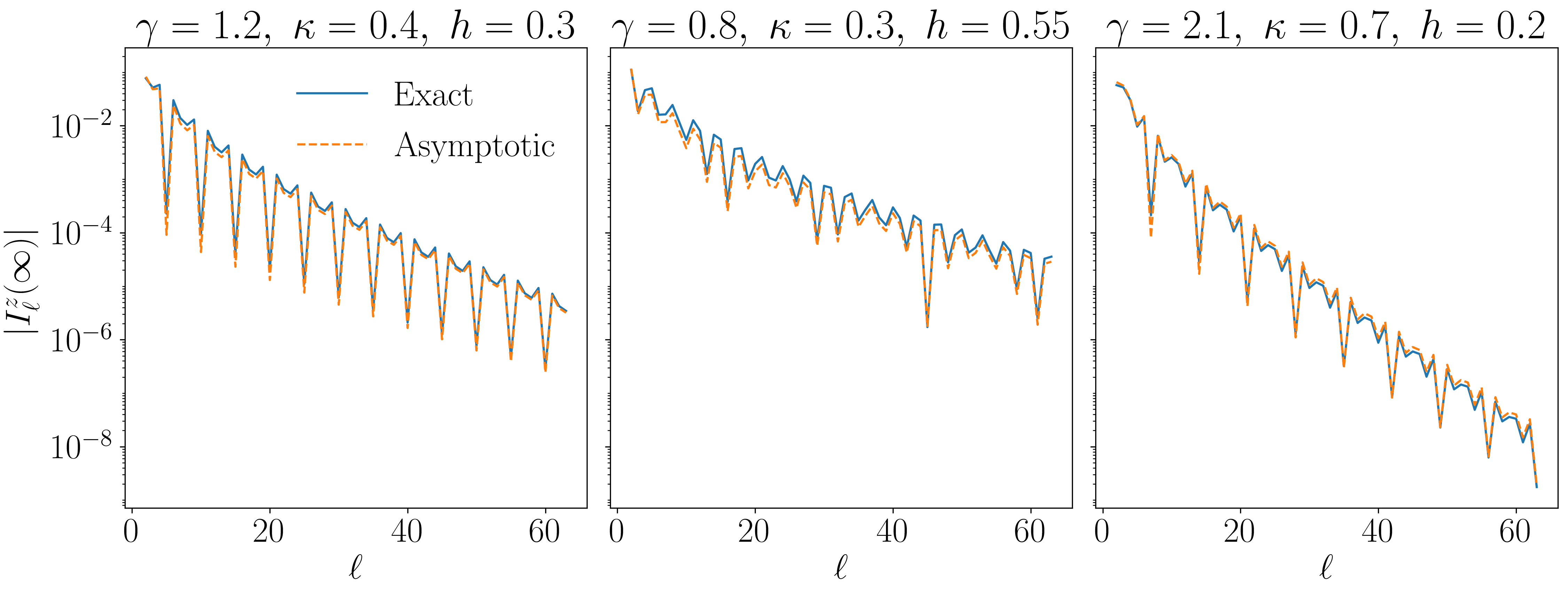}
    \caption{\label{fig:Iz} Comparison between the exact expression for $I^z_l(\infty)$ and the asymptotic behaviour Eq.~\eqref{eq:IZEXACT} within the imaginary gapless phase. As predicted in the text, the spatial correlations decay exponentially. In this case, the leading order Eq.~\eqref{eq:IZEXACT} fully encode the qualitative behavior of the function, but in regions where the singularities compete, next-leading orders need to be carefully extracted. }
\end{figure}

From these expressions we can compute the stationary $u_k^\infty \equiv u_k(\infty)$ and $v_k^\infty \equiv v_k(\infty)$
\begin{align}
   C u_k^\infty & \simeq - 2 i \tilde\gamma - 2\frac{k-k^*}{|k-k^*|}  \sqrt{\kappa^2(1-h^2)-\tilde\gamma^2} + 2 (k-k^*) \sqrt{1-h^2} \nonumber\\&\qquad \qquad + 2 |k-k^*|\frac{\sqrt{1-h^2}}{\sqrt{\kappa^2(1-h^2)-\tilde\gamma^2}}\left( {\kappa^2 h } + i {\tilde\gamma}\right),\\
   C v_k^\infty & \simeq - 2 \kappa \left(\sqrt{1-h^2}-h (k-k^*)\right).
\end{align}
We need lastly to compute three objects, which are $u_k^\infty \overline{v^\infty}_k$, $|u^\infty_k|^2$ and $|v^\infty_k|^2$, given respectively by
\begin{align}
    |C|^2 \mathrm{Re}(u^\infty_k \overline{v^\infty}_k) &= 4 \kappa \sqrt{1-h^2}\frac{k-k^*}{|k-k^*|}\sqrt{\kappa^2(1-h^2)-\tilde\gamma^2} - 4 \kappa (k-k^*) (1-h^2)\nonumber\\ &\qquad\quad + 4 \kappa  h |k-k^*|\left(\sqrt{\kappa^2(1-h^2)-\tilde\gamma^2} - 4 \kappa^3 h  \frac{1-h^2}{\sqrt{\kappa^2(1-h^2)-\tilde\gamma^2}}\right),\label{eq:1111}\\
    |C|^2 \mathrm{Im}(u^\infty_k \overline{v^\infty}_k) &= 4\tilde\gamma \kappa \left(\sqrt{1-h^2} - h (k-k^*) - |k-k^*| \frac{\sqrt{1-h^2}}{\sqrt{\kappa^2(1-h^2) - \tilde\gamma^2}}\right), \label{eq:1112}\\
    |C|^2 |v^\infty_k|^2& = 4\kappa^2 (1-h^2) - 8 \kappa^2 h \sqrt{1-h^2} (k-k^*) \label{eq:1113} \\
    |C|^2 |u^\infty_k|^2& = |C|^2 |v^\infty_k|^2 - 8 |k-k^*| \sqrt{1-h^2} \sqrt{\kappa^2 (1-h^2) -\tilde\gamma^2}  - 8|k-k^*|\tilde\gamma^2 \frac{\sqrt{1-h^2}}{\sqrt{\kappa^2(1-h^2)-\tilde\gamma^2}}.\label{eq:1114}
\end{align}
These are the ingredients to evaluate the integrals Eq.~\eqref{eq:integrals}, which we compute at leading order in $1/l$. We have
\begin{align}
    I^x_l(\infty) &\simeq \frac{1}{2\pi}\int_{-\pi}^{\pi} dk e^{-i k l} 2\frac{k-k^*}{|k-k^*|}\sqrt{1-\frac{\tilde\gamma^2}{\kappa^2(1-h^2)}} = \frac{4}{\pi}\sqrt{1-\frac{\tilde\gamma^2}{\kappa^2(1-h^2)}}\frac{\cos(k^* l)}{l},\label{eq:IXEXACT}\\
    I^y_l(\infty) &\simeq -\frac{1}{2\pi}\int_{-\pi}^{\pi} dk e^{-i k l} |k-k^*|\frac{1}{\kappa^2 \sqrt{1-h^2}}\left(\sqrt{\kappa^2(1-h^2)-\tilde\gamma^2} + \frac{\tilde\gamma^2}{\sqrt{\kappa^2(1-h^2)-\tilde\gamma^2}}\right)\nonumber\\
    &\quad =-\frac{4}{\pi} \frac{1}{\kappa^2 \sqrt{1-h^2}}\left(\sqrt{\kappa^2(1-h^2)-\tilde\gamma^2} + \frac{\tilde\gamma^2}{\sqrt{\kappa^2(1-h^2)-\tilde\gamma^2}}\right) \frac{\cos(k^* l)}{l^2}\label{eq:IYEXACT}
\end{align}
\begin{figure}[t]
    \centering
    \includegraphics[width=.8\columnwidth]{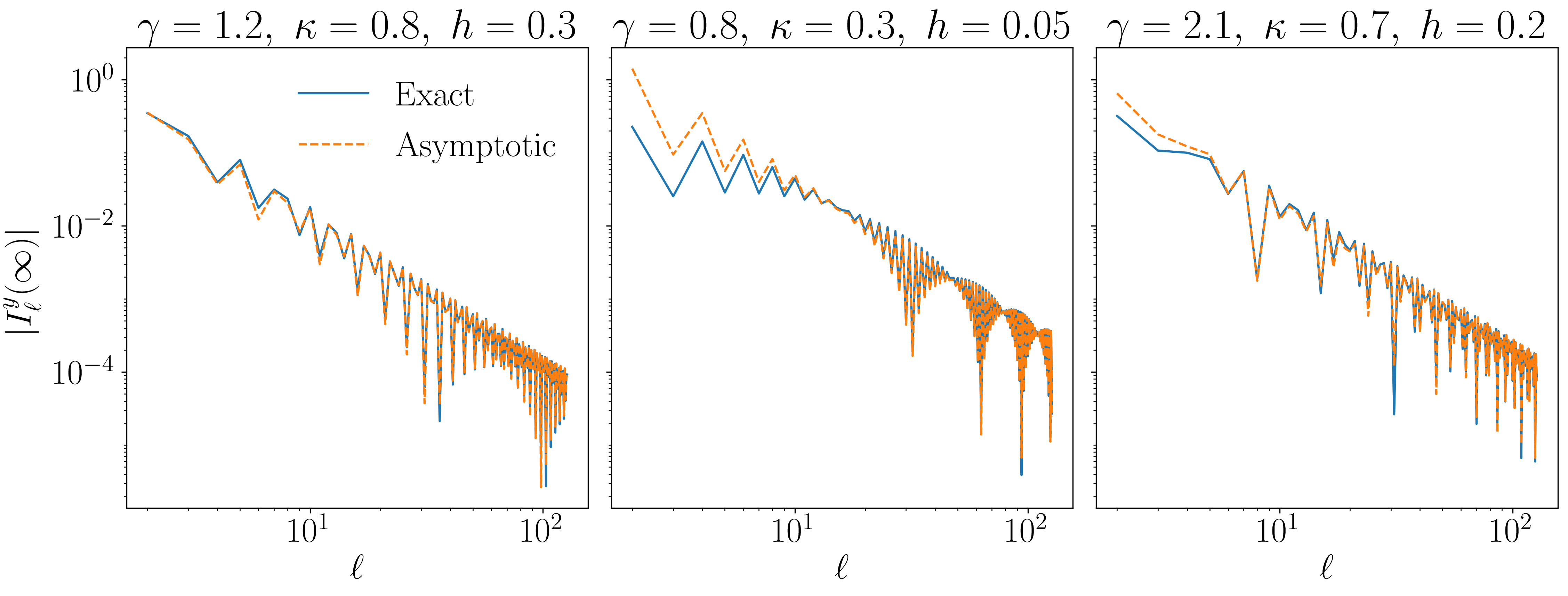}
    \caption{\label{fig:Iy} Comparison between the exact expression for $I^y_l(\infty)$ and the asymptotic behaviour Eq.~\eqref{eq:IYEXACT} within the imaginary gapless phase. Similarly to the $I^x_l(\infty)$ case, the agreement is quantitative already for modest distance $l$. }
\end{figure}
Up to now, the integrals we had to deal with had the leading singularity at $k\simeq k^*$. This does not hold for $\mathrm{Im}(u_k^\infty \overline{v^\infty_k})$, where the value $k\simeq k^*$ gives an analytic contribution. A simple argument is given by writing $\Gamma_k = 4 \tilde\gamma (\cos(k) + h) / E_k$, expanding both numerator in series of $(k-k^*)$ and noticing that all the integrals contribute with derivatives of the delta-functions $\delta(l)$. 
Hence, we can already conclude the leading order of the integral is not algebraic. To compute the asymptotic behavior, we resort to the complex contour integral induced by the transformation $z = i e^{ik}$ around the unit circle. The singular behavior is dictated by the roots of $\Lambda_z\equiv \Lambda_{k(z)}$, given by
\begin{equation}
    z^{p_1}_{p_2} = \frac{p_1(i h- \tilde\gamma) - p_2\sqrt{1-h^2 - 2 i h \tilde\gamma + \tilde\gamma^2 - \kappa^2}}{\kappa - p_1},\qquad p_1,p_2 = \pm 1.
\end{equation}
After a lengthy but straightforward computation we find at leading order
\begin{equation}
    I^z_l(\infty) = \frac{2}{\pi}\frac{1}{l^{3/2}}  \mathrm{Im}(i z^\star)^l,\qquad z^\star = \mathrm{argmax}_{z^{p_1}_{p_2}:|z^{p_1}_{p_2}|<1} (-1/\log |z^{p_1}_{p_2}|),\label{eq:IZEXACT}
\end{equation}
where $z^\star$ is the root with larger weight in the integral. We test the predictions for $|I^\alpha_l(\infty)|$, and report in Fig.~\ref{fig:Ix}, Fig.~\ref{fig:Iy} and Fig.~\ref{fig:Iz} some of these comparisons.
(We remark, without detailing, that similar contour integral computations arise also for the gapped phase, where no gapless modes exists). 

We conclude this section by noting that the stationary state spin-spin correlation function is, at leading order, in the gapless phase
\begin{equation}
    |C^\mathrm{zz}_x(\infty)| \simeq \left(\frac{4}{\pi}\right)^2 \left(1-\frac{\tilde\gamma^2}{\kappa^2(1-h^2)}\right) \frac{\cos^2(k^* x)}{|x|^2},
\end{equation}
whereas is exponentially suppressed in the gapped regime for distances greater than the correlation length $|x|>\xi_\mathrm{corr}$.

\section{S4 Dynamics of entanglement entropy}
Due to the Gaussianity of the state, the entanglement entropy is characterized by the two-point fermionic functions~\cite{jin2005entanglement,peschel2004on}. Throughout this section we consider a bipartition $X\cup X_c$ where $X$ is a connected interval of length $\ell$ and $X_c$ is its complementary. For convenience we introduce the Majorana fermions
\begin{align}
	\hat{a}_{2l-1} = \hat{c}^\dagger_l + \hat{c}_l = \hat{A}_l,
	\qquad \hat{a}_{2l}  = -i (\hat{c}^\dagger_l - \hat{c}_l) = -i \hat{B}_l.
\end{align}
for which the correlation function is given by
\begin{align}
	\langle \Psi_t|\hat{a}_m\hat{a}_n|\Psi_t\rangle = \delta_{mn} - \Gamma^\ell_{mn}
\end{align}
where the matrix $\Gamma^\ell$ reads
\begin{align}
	\Gamma^\ell= \begin{pmatrix}
		\Pi_0 & \Pi_{-1} & \cdots &\Pi_{1-{\ell}}\\
		\Pi_1 & \Pi_{0} & \cdots & \vdots\\
		\vdots & \vdots & \ddots & \vdots \\
		\Pi_{\ell-1} & \cdots & \cdots &\Pi_{0}
	\end{pmatrix},\qquad \Pi_l = \frac{1}{2\pi} \int_{-\pi}^{\pi} dk e^{-i k l} \vec{\Pi}(k)\cdot \vec{\sigma}
\end{align}
where $\vec{\Pi}(k)=(\Pi_x(k),\Pi_y(k),\Pi_z(k))$ are the matrix introduced in the previous section.
The entanglement entropy is then given by
\begin{equation}
    S_t(\ell) = -\mathrm{tr}\left(\frac{1-\Gamma^\ell}{2}\ln \frac{1-\Gamma^\ell}{2} \right)\label{eq:defent}
\end{equation}

Because $||\vec{\Pi(k)}||^2 = 1$ for all $k$, using the Szeg\"o theorem on the symbol $\Pi(k) \equiv \vec{\Pi}(k)\cdot \vec{\sigma}$ it follows the absence of a volume law. We introduce the functions $e(y,x) = -(x+y)/2\ln[(x+y)/2] - (y-x)/2\ln[(y-x)/2]$, $D^\ell(\lambda) = \lambda - \Gamma^\ell$ and $\mathbb{P}(k) = \lambda - \Pi(k)$. Using Cauchy theorem, the sum Eq.~\eqref{eq:defent} is rewritten as a complex integral over the zeros of $D^\ell(\lambda)$ as
\begin{equation}
    S_t(\ell) = \frac{1}{4 \pi i} \oint_\mathcal{C} d\lambda e(1+0^+,\lambda)\frac{d}{d\lambda}D^\ell(\lambda)\;, \label{eq:ziozio}
\end{equation}
with $\mathcal{C}$ a line surrounding all the zeros of $D^\ell$. 
In the limit $t \to \infty$, Szeg\"o theorem~\cite{jin2005entanglement} express the leading order of Eq.~\eqref{eq:ziozio} as 
\begin{equation}
    S_\infty(\ell) \simeq \frac{\ell}{8 \pi^2 i} \oint_\mathcal{C} d\lambda e(1+0^+,\lambda) \int_0^{2\pi} dk \frac{d}{d\lambda} \ln \det \mathbb{P}(k)+ \mathcal{O}(\ln \ell).\label{eq:ziocane}
\end{equation}
A simple computation shows that the poles are around $\lambda \pm 1$ (since $\det\Pi(k) = 1$ for all $t$ and $k$), for which $e(1+0^+,\lambda = \pm 1)=0$. Therefore the linear term in $\ell$ vanishes, \textit{i.e.} showing the absence of any volume law. 
A similar discussion extends for the scaling limit $t\sim \ell \gg1$. Using the series expansion in Ref.~\cite{fagotti2008evolution} for Eq.~\eqref{eq:defent}, one can show that the term proportional to $t$ vanishes identically. 

Therefore, it is the subleading term in Eq.~\eqref{eq:ziocane} which encodes the entanglement properties, both of the time evolution and of the stationary state. Since the calculation is harder within the time-evolution, we hereby compute the entanglement of the stationary state, which is already meaningful to characterize the dynamical phases of the system. In this limit, we have $\Pi_l \simeq \vec{I}_l(\infty)\cdot \vec{\sigma}$. 
In full analogy to other systems considered in the literatures ~\cite{yates2018central,ares2015entanglement,ares2019sublogarithmic,fraenkel2021entanglement}, the subleading term is logarithmic in $\ell$ and can be computed following the guidelines in Ref.~\cite{peschel2004on}.

\begin{figure}[h!]
    \centering
    \includegraphics[width=0.6\columnwidth]{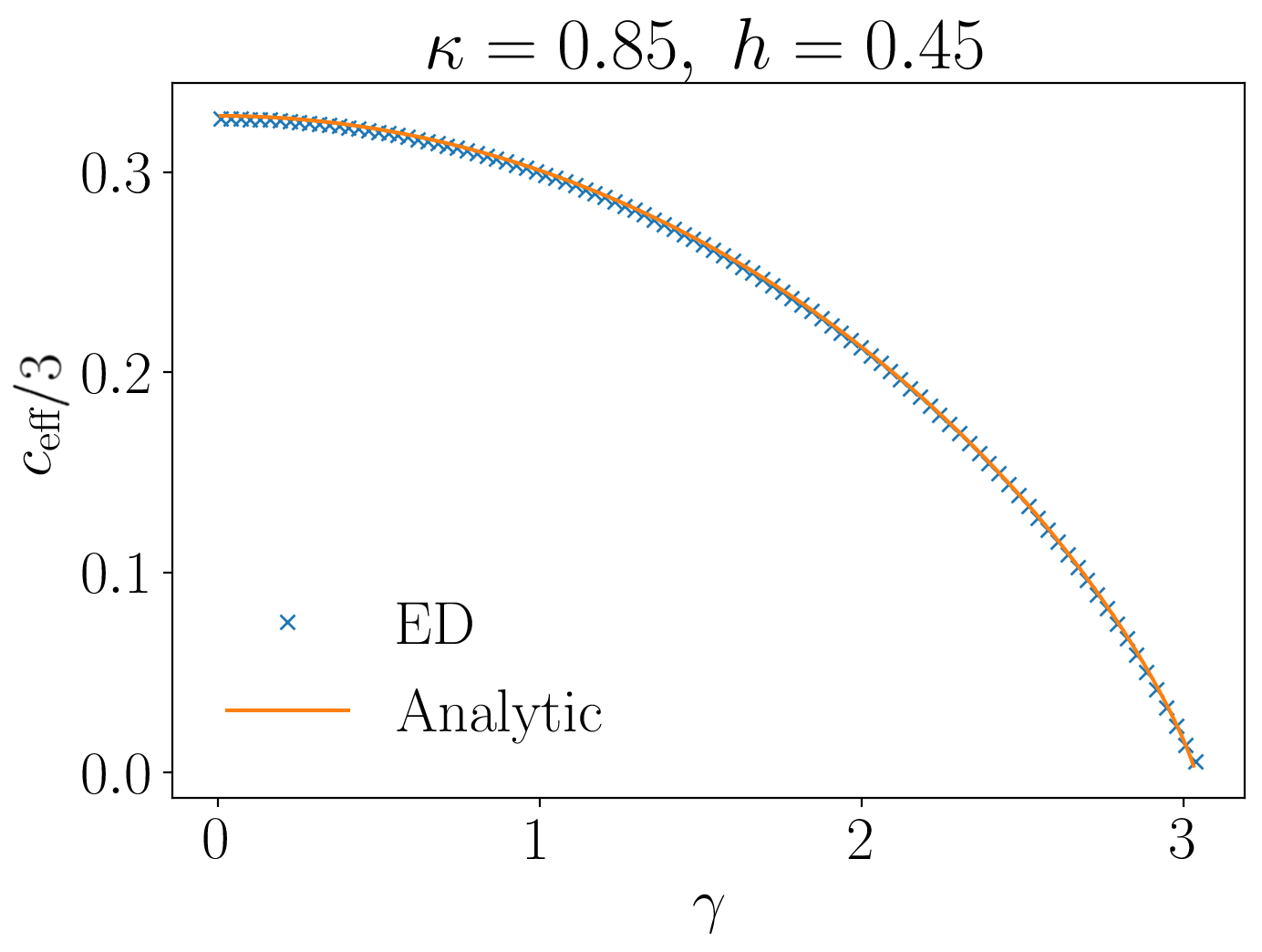}
    \caption{\label{fig:cefftest} Comparison between the formula Eq.~\eqref{eq:ceff} )in orange), and the \textit{ab-initio} numerics for a choice of parameters $\kappa$, $h$.}
\end{figure}

Using the formulae Eqs.~\eqref{eq:1111}-\eqref{eq:1114}, at leading order in $(k-k^*)$ the correlation matrix is given by 
\begin{equation}
    \Gamma_{ij}^X(\infty) = \int_{-\pi}^\pi \frac{dk}{2\pi} e^{-i k (i-j)}\tilde{\Pi}(k),\qquad \tilde{\Pi}(k) \equiv \begin{pmatrix}
    -\beta  & \sqrt{1-\beta^2} \frac{k-k^*}{|k-k^*|}\\
    \sqrt{1-\beta^2} \frac{k-k^*}{|k-k^*|}  & \beta,
    \end{pmatrix}
\end{equation}
where $\beta=\tilde\gamma/(\kappa\sqrt{1-h^2})$. Integrating we have
\begin{equation}
    \Gamma_{ij}^X = -\beta \delta_{ij}\sigma^z + \sqrt{1-\beta^2} \left(\delta_{ij} - i(1-\delta_{ij})\frac{e^{-i k^*(i-j)}}{\pi (i-j)}\right)\sigma^x.
\end{equation}
To evaluate Eq.~\eqref{eq:defent} we need the spectrum of $\Gamma^\ell$
\begin{equation}
    \sum_{j} \Gamma^\ell_{ij} \begin{pmatrix} \omega_j \\ \Omega_j 
    \end{pmatrix} = \lambda \begin{pmatrix} \omega_i \\ \Omega_i 
    \end{pmatrix}.
\end{equation}
It is convenient to absorb the phase $\omega_j, \Omega_j \mapsto e^{i k^* j}\omega_j, \Omega_j$, which does not affect the eigenvalues of $\Gamma^\ell$. We have
\begin{align}
    i \sum_j D^+_{ij} \omega_j &= \lambda \Omega_i \nonumber\\
    i \sum_j D^-_{ij} \Omega_j &= \lambda \omega_i \label{eq:corrmatgamma}
\end{align}
where $D^{\pm}_{ij} \equiv (\sqrt{1-\beta^2} \pm i\beta) \delta_{ij} + \sqrt{1-\beta^2}/(\pi (i-j))$. Then
\begin{equation}
    \sum_{lj} D^-_{il}D^+_{lj} \omega_j = -\lambda^2 \omega_i.~\label{eq:eigprob}
\end{equation}
The eigenproblem Eq.~\eqref{eq:eigprob} can be solved easily following Ref.[Peschel,altri], and here we add the details for completeness. We define $K_{il} \equiv \sum_j -D^-_{ij}D_{jl}^+/4$
\begin{equation}
    4 \sum_{l| l\neq i} K_{il} \omega_l = (\lambda^2 - 4 K_{ii})\omega_i.
\end{equation}
From the previous expression Eq.~\eqref{eq:corrmatgamma}, $K_{ii} = 1/4$, whereas for $i\neq l$
\begin{equation}
    K_{il} = -\frac{\sqrt{1-\beta^2}}{2\pi^2 (i-l)}\sum_j \left(\frac{1}{2(i-j)} + \frac{1}{2(j-l)}\right).
\end{equation}
Taking the continuous limit of the above kernel we have
\begin{align}
    \frac{1}{2} \int_{1/\ell}^{2-1/\ell} dx' K(x,x') \omega(x') &= \left(\frac{\lambda^2}{4} - K_{ii}\right)\omega(x),\\
    K(x,x') &=  -\frac{\sqrt{1-\beta^2}}{2\pi^2 (x-x')}\left[\ln \left(\frac{x}{2-x}\right) - \ln \left(\frac{x'}{2-x'}\right)\right].
\end{align}
The solution of the above integral equation can be obtained after the change of variable
\begin{equation}
    u(x) \equiv \frac{1}{2}\ln \left(\frac{x}{2-x}\right),\qquad \chi(u) = \omega(u)/\cosh(u)
\end{equation}
which gives
\begin{equation}
    -\frac{1}{2\pi^2}\int_{-\infty}^{\infty} du' \frac{u-u'}{\sinh(u-u')}\chi(u') = \left(\frac{\lambda^2}4 - K_{ii}\right) \chi(u).
\end{equation}
Fourier transforming the above problem over the frequency domain $q$, the convolution of functions in $u$ is transformed in a product of functions in $q$. Since
\begin{equation}
    -\frac{1}{2\pi^2}  \int_{-\infty}^{\infty} du' \frac{u'}{\sinh( u')}e^{-i q u'} = -\frac{1}{4 \cosh^2(\pi q/2)},
\end{equation}
the eigenvalues are given by 
\begin{equation}
    \lambda_q = \pm\sqrt{\frac{\sinh^2(\pi q/2) +\beta^2}{\cosh^2(\pi q /2)}}\label{eq:eigvals}.
\end{equation}
The quantization of $\lambda_q$ is fixed by the boundary condition $\omega(1) = \pm \omega(\ell)$ which is a consequence of the parity. For large $\ell$ this is $q_n = n \pi/\log \ell$. Lastly
\begin{equation}
    S(X) = -\sum_{n} \left(\frac{1-\lambda_n}{2}\ln\frac{1-\lambda_n}{2} + \frac{1+\lambda_n}{2}\ln\frac{1+\lambda_n}{2} \right).\label{eq:maronn}
\end{equation}
The remaining is a lengthy but straightforward computation, which require the change of variable $n=n(\lambda)$ and the conversion of Eq.~\eqref{eq:maronn} to an integral. This variable change is explicitly given by 
\begin{equation}
    n = -i \frac{2}{\pi^2}(\ln \ell) \mathrm{arccsc} \left(\frac{ 2e^{\lambda/2}}{\sqrt{\beta^2(1+e^\lambda)^2 - (1-e^\lambda)^2}}\right)\;,
\end{equation}
which can be obtained through the chain rule $n=(2\ln \ell)/(\pi^2) x$ and $x=x(\lambda)$. Inserting this expression in Eq.~\eqref{eq:maronn} and after trivial algebra we get the final expression for the entanglement entropy. For convenience we report the final expression also in this Supplemental Material
\begin{align}
    S(X) &= \frac{1}{3}c_\mathrm{eff} \ln \ell + O(1),\qquad \beta\equiv \frac{\tilde\gamma}{\kappa\sqrt{1-h^2}}\\
    c_\mathrm{eff} & = \frac{12}{\pi^2}\mathrm{Re}\int_0^1 d\lambda \Upsilon(\lambda) \frac{\lambda}{(1-\lambda^2)}\frac{\sqrt{\beta^2-1}}{\sqrt{\beta^2-\lambda^2}}\label{eq:ceff}
\end{align}
where in the above equation we have introduced the entropy function
\begin{equation}
    \Upsilon(x) = -\frac{1+x}{2}\ln\left(\frac{1+x}{2}\right)- \frac{1-x}{2}\ln\left(\frac{1-x}{2}\right).
\end{equation}
A comparison between the analytic expression Eq.~\eqref{eq:ceff} and the exact numerics is given in Fig.~\ref{fig:cefftest}.

\end{document}